\documentclass[onecollarge,natbib]{svjour2}
\bibpunct{[}{]}{;}{n}{}{,} % to get "[numbered]" references from natbib
\smartqed  % flush right qed marks, e.g. at end of proof
\journalname{Few Body Systems}
\usepackage{color}
\definecolor{mBlue}{rgb}{0,0,1}
 
\definecolor{mRed}{rgb}{1,0,0}
 \newcommand{\mRed}[1]{\textcolor{mRed}{#1}}
\definecolor{mGreen}{rgb}{0,1,0}

\usepackage{srcltx}

 \usepackage[dvips,final]{graphicx}
  \usepackage{amssymb,amsmath,epsfig,bm,pifont}
   \graphicspath{{./figs/}}

\newcommand{\va}[1]{\langle{#1}\rangle}                            %%%%%%%%%
\newcommand{\gev}[1]{\relax\ifmmode{\text{GeV}^{#1}}               %%%%%%%%%
                     \else{{GeV}$^{#1}${ }}\fi}                    %%%%%%%%%

\begin{document}

\title{Rho meson distribution amplitudes from QCD sum rules with nonlocal condensates}


\titlerunning{$\rho$ meson DAs}        % if too long for running head

\author{A.~V.~Pimikov\footnote{
	Presented by the first author at the Light-Cone Conference 2013, May 20-24, 2013, Skiathos, Greece.}
	\and S.~V.~Mikhailov \and N.~G.~Stefanis
        }

\institute{A.~V.~Pimikov \at
                Departamento de F\'{\i}sica Te\'orica -IFIC,
                Universidad de Valencia-CSIC, E-46100 Burjassot
                (Valencia), Spain
                Bogoliubov Laboratory of Theoretical Physics, JINR,
                141980 Dubna, Russia\\
                \email{alexandr.pimikov@uv.es} \and
           S.~V.~Mikhailov \at
                Bogoliubov Laboratory of Theoretical Physics, JINR,
                141980 Dubna, Russia\\
                \email{mikhs@theor.jinr.ru} \and
           N.~G.~Stefanis \at
                Institut f\"{u}r Theoretische Physik II,
                Ruhr-Universit\"{a}t Bochum,
                D-44780 Bochum, Germany\\
                \email{stefanis@tp2.ruhr-uni-bochum.de}
           }

\date{Received: date / Accepted: date}

\maketitle

\begin{abstract}
The leading-twist distribution amplitude for the longitudinal
rho-meson was studied using QCD Sum Rules with nonlocal condensates
and a spectral density which includes next-to-leading order radiative
corrections.
The obtained profile is compared with results from standard
QCD sum rules, lattice QCD, holographic QCD, a light-front quark model,
and the instanton liquid model.
Preliminary estimates for the first two moments of the transverse
$\rho$-meson distribution amplitude are also given.

\keywords{Rho-meson properties
\and Meson Distribution Amplitudes
\and QCD Sum Rules}
\end{abstract}

\section{Introduction}
\label{sec:intro}

The knowledge of the partonic $\rho$-meson structure is very important
for the theoretical description of different hard exclusive processes,
such as the B-meson decays
($B\to \rho l\nu$, $\overline B^0\to \rho^0\gamma$)~\cite{BB97,AhSa12},
the $\rho$-meson electro-photoproduction~\cite{FS12},
and the determination of the Cabibbo--Kobayashi--Maskawa matrix element
$|V_{ub}|$~\cite{BB97}.
In this report, we focus on the construction of the leading-twist
distribution amplitude (DA), $\varphi_\rho^\text{L}(x,\mu^2)$, for the
longitudinal $\rho$ meson, defined by
\begin{equation}
  \va{0\!\mid\! \bar
  d(z) \gamma_\mu u(0)\!\mid\! \rho(p,\lambda)}\Big|_{z^2=0}
=
  f_\rho p_\mu\,
  \!\!\int_{0}^{1}\!\!\!\! dx\ e^{ix (z \cdot p)}\,
  \varphi_\rho^\text{L}(x,\mu^2) \, .
\label{def:rhoL}
\end{equation}
%Eq (1)
On the other hand, the transverse $\rho$-meson DA,
$\varphi_\rho^\text{T}(x,\mu^2)$, is given by
\begin{equation}
  \va{0\!\mid\! \bar d(z) \sigma_{\mu\nu}u(0)\!\mid\!\rho(p,\lambda)}\Big|_{z^2=0}
=
  i f_\rho^\text{T}(\varepsilon^{(\lambda)}_\mu p_\nu-\varepsilon^{(\lambda)}_\nu p_\mu)\,
  \!\!\int_{0}^{1}\!\!\!\! dx\ e^{ix (z \cdot p)}
  \varphi_\rho^\text{T}(x,\mu^2) \, ,
\label{def:rhoT}
\end{equation}
%Eq (2)
where $p_{\nu}$ and $\varepsilon^{(\lambda)}_\mu$
are the momentum and polarization vector of the $\rho$  meson,
respectively, and $\mu^2$ is the normalization point.
A fuller description of this DA will be given in a separate
publication.
Here, we will only present preliminary estimates for its first two
moments.
The evolution with the factorization scale $\mu^2$ of the
$\varphi_\rho^\text{L}(x,\mu^2)$ and $\varphi_\pi(x,\mu^2)$
DAs is taken into account by means of the
Efremov--Radyushkin--Brodsky--Lepage
evolution equation~\cite{ER80,LB80}.
Note that the next-to-leading order (NLO) evolution of
$\varphi_{\rho}^{\rm L}(x, \mu^2)$
resembles that of $\varphi_{\pi}(x, \mu^2)$ ---
see, e.g., \cite{MR86ev}.

The $\rho_{(\rm L,T)}$-meson DAs (longitudinal and transverse)
were first constructed within the standard QCD SR approach
\cite{CZ84,BB96}, which employs local condensates.
Their determination within the framework of QCD SRs with nonlocal
condensates (NLC) was considered later in \cite{BM98,BM00,BM01}.
Besides, the $\rho$-meson DAs were studied in other nonperturbative
approaches as well, e.g.,
lattice QCD~\cite{Lat10},
AdS/QCD holography~\cite{TB08,VSBG09,AhSa12,FS12},
the light-front quark model~\cite{CJ07},
and the instanton liquid model~\cite{D06}.
In this note we improve the previous works in \cite{BM98,BM00,BM01}
on the $\rho$-meson longitudinal and transverse DAs in
the following respects:
(i) we use a corrected expression for the NLO spectral density
in the SR for the $\rho_{\rm L}$ DA,
(ii) we express the $\rho_{\rm L}$ DA in terms of Gegenbauer
coefficients, and
(iii) we calculate and use the correct nonperturbative term for the
theoretical part of the SR for the $\rho_{\rm T}$ DA.
We utilize the latter improvement to derive some preliminary
estimates for the decay constant and the two lowest-order
moments of the $\rho_{\rm T}$-meson DA.

\section{Distribution amplitudes for the $\rho$-meson from QCD sum rules
         with nonlocal condensates}
\label{sec:SRrhoL}
The derivation of the $\rho$-meson longitudinal DA is based on the
correlator of two vector currents that leads to the following
sum rule for the $\rho$-DA $\varphi_\rho^{\rm L}(x)$:
\begin{subequations}
\label{eq:SRrhoL}
\begin{eqnarray}
  \!\!\!\!f_{\rho}^2\,\varphi^\text{L}_\rho(x)e^{-m_{\rho}^2/M^2}
  + f_{\rho'}^2\,\varphi^\text{L}_{\rho'}\!(x)\, e^{-m_{\rho'}^2/M^2}
\!=\!\!
  \int\limits_{\!\!0}^{~~s_0}%{\infty}
  \!\!\rho_\text{pert}\left(s,x\right)
  e^{-s/M^2}\!\!ds
  + \Phi_\rho(x,M^2) \, ,&&  \\
  \!\! \Phi_{\rho(\pi)}(x,M^2)
=
  \mp \Phi_\text{4Q}(x,M^2)\!
  + \Phi_{\bar q Aq}(x,M^2)\!
  + \Phi_\text{V}(x,M^2)\! + \Phi_\text{G}(x,M^2) \, .
\label{eq:Phi}
&&
\label{eq:SRrhoLconden}
\end{eqnarray}
\end{subequations}
%Eq (3a), (3b)
Here, $M^2$ is the Borel parameter, $s_0$ denotes the threshold
value, and the nonperturbative contribution $\Phi_\rho$ to the
operator product expansion (OPE) contains the following terms:
$\Phi_{\text{4Q}}$ (four-quark condensate),
$\Phi_{\bar q Aq}$ (quark-gluon-quark condensate),
$\Phi_\text{V}$ (vector quark condensate),
$\Phi_\text{G}$ (gluon condensate).
Note that the r.h.s. of the analogous sum rule for the pion DA receives
the same nonperturbative contributions, but has a positive
sign in front of the dominant term $\Phi_{\text{4Q}}$
determined by the scalar quark condensate
\cite{BM98}.\footnote{The generic form of the sum rule \cite{BM00,BM01}
for the transverse $\rho_{\rm T}$ DA, used in our analysis, is similar
to (3a) (apart from the appearance on the l.h.s. of the masses of
$\rho$ and $\rho'$).}
We will discuss later the implications of this sign difference.
All these nonperturbative contributions contain nonlocal
condensates \cite{MR89,MR92}, meaning that the quarks (gluons) in the
vacuum can be correlated within distances of the order of
$\Lambda=1/\lambda_q ~(1/\lambda_G)$.
The value of this length scale is fixed by the nonzero average
virtuality of the quark fields in the vacuum condensate,
$\lambda_q^2\equiv\va{\bar qD^2q}/\va{\bar qq}$,
which is the only mass-scale setting parameter used to
parameterize the nonlocality of the nonperturbative vacuum.
A natural and simple ansatz for the spatial ($z^2 < 0$) behavior of
the scalar quark condensate is given by a Gaussian model, e.g.,
$\va{\bar{q}(z)q(0)} \sim \exp[z^2 \lambda_q^2/8]$ \cite{MR89,MR92}
with
$\lambda_q^2 = 0.4\pm 0.05 \text{GeV}^2$ \cite{BM02,BMS03}.

Due to the nonlocality, the behavior of
$\Phi_{\rho(\pi)}(x,M^2)$ becomes less singular with respect to
$x$.
For instance, $\Phi_\text{4Q}(x,M^2)$ in Eq.\ (\ref{eq:SRrhoLconden})
shows a linear behavior in the vicinity of the endpoints --- in
contrast to the singular $\delta(x)$-behavior in the standard,
i.e., local, approach,
$$
  \Phi_\text{4Q}\sim x\theta(\Delta-x)
~\stackrel{\lambda^2_q \rightarrow 0}{\longrightarrow}~
  \Phi^\text{local}_\text{4Q}\sim\delta(x) \, ,
$$
where $\Delta=\lambda_q^2/(2M^2)\in[0.1,0.3]$.
In order to reduce the model dependence, introduced via the
nonlocality ansatz, we study the sum rules for the integral
characteristics of the meson DAs, such as the moments
$\va{\xi^{N}} \equiv \int_{0}^{1}\!\! dx\varphi(x)(1-2x)^{N}$\,
and the inverse moment
$\langle x^{-1}\rangle\equiv \int_{0}^{1}\!\! dx\varphi(x)/x$.
In fact, the nonlocal approach gives the opportunity to
study the slope $\varphi'(x\to 0)$~\cite{MPS10} and the inverse
moment
$\langle x^{-1}\rangle$ \cite{BMS01}
of the meson DAs.
These characteristics are inaccessible when one uses local condensates
because of the appearance of non-integrable singularities.

Following \cite{BMS01}, we construct
$\varphi_\rho^\text{L}(x,\mu^2)$
in terms of the first few Gegenbauer coefficients which we
extract from the moments of the $\rho_{\rm L}$ DA, viz.,
$\va{\xi^{2N}}$
with $N=0,1,\ldots ,5$.
The moments themselves are determined by evaluating the sum
rules in Eq.\ (\ref{eq:SRrhoL}).
The obtained results are displayed in Tab.\ \ref{tab:rhoL}.
Still higher Gegenbauer coefficients can be set equal to zero,
i.e., $a_{\tiny n\geq 6}=0$ --- a reasonable approximation in view of
the closeness of the higher moments to their asymptotic values, cf.\
the entries in the first and the fourth row of
Tab.\ \ref{tab:rhoL}.

%%%%%%%%%%%%%%%%%%%%%%%%%%%%%%%%%%%%%%%%%%%%%%%%%%%%%%%%%%%%%%%%%%%%%%% Table 1
\begin{table*}[th!]
\caption{Results for the moments of various $\rho$-meson DAs
derived in our present analysis employing QCD sum rules with
nonlocal condensates --- cf.\ Eq.\ (\ref{eq:SRrhoL}).
The normalization scale is taken to be
$\mu^2 \approx 1$~GeV$^2$.
The analogous results for the asymptotic DA and
the pion DA from Ref.\ \cite{BMS01}
are also given for comparison.
\label{tab:rhoL}}
\centering
\begin{tabular}{cccccccc}
\hline\noalign{\smallskip}
Meson DA            & $f_{M}$   & $\langle\xi^2\rangle$
                            & $\langle\xi^4\rangle$
                                                        & $\langle\xi^6\rangle$
                                                                    & $\langle\xi^8\rangle$
                                                                                & $\langle\xi^{10}\rangle$
                                                                                           & $\langle x^{-1}\rangle_\text{SR}$
\\[3pt] \tableheadseprule\noalign{\smallskip}
Asy                 & 1         & 0.2       & 0.0857    & 0.0476    & 0.030     & 0.0209   & 3        \\
$\pi$~\cite{BMS01}  & 0.137(8)  & 0.266(20) & 0.115(11) & 0.060(7)  & 0.036(5)  & 0.025(4) & 3.35(30)  \\
$\rho_L$~\cite{BM98}& 0.201(5)  & 0.227(7)  & 0.095(5)  & 0.051(4)  & 0.030(2)  & 0.020(5) & 3.1(1)     \\
$\rho_L$  (here)    & 0.21(2)   & 0.216(21) & 0.089(9)  & 0.048(5)  & 0.030(3)  & 0.022(2) & 3.16(30)    \\
$\rho_L'$ (here)    & 0.181(26) & 0.273(65) & 0.175(30) & 0.120(17) & 0.083(12) & 0.058(8) & 4(1)         \\
$\rho_T$  (here)    & 0.169(15) & 0.107(11) & 0.022(2)  & --        & --        & --       & --           \\
\noalign{\smallskip}\hline
\end{tabular}
\end{table*}
%%%%%%%%%%%%%%%%%%%%%%%%%%%%%%%%%%%%%%%%%%%%%%%%%%%%%%%%%%%%%%%%%%%%%%%

The evaluation of the SR in Eq.\ (\ref{eq:SRrhoL}) from the
numerical viewpoint is carried out at the scale
$\mu^2 \approx 1$~GeV$^2$ in the following way.
For the masses of the $\rho$-mesons we use their physical
values from \cite{PDG12}, i.e.,
$m_\rho = 0.775$~GeV, $m_{\rho'} = 1.496$~GeV, and
$m_{\rho''} = 1.72$~GeV.
The size of the nonlocality parameter $\lambda^2_q$
was determined in the analysis of the pion DA, carried out in
\cite{BMS01,BMS02,BM02} and was found to be $\lambda_q^2=0.4$~GeV$^2$.
(We ignore here for simplicity its uncertainties, mentioned above.)
The coupling constant has the value
$\alpha_s(1~\text{GeV}^2)=0.56$,
whereas the quark condensate and the gluon
condensate are given by
$\alpha_s\va{\bar qq}^2=0.000183$~GeV$^6$ and
$\alpha_s\va{GG}/\pi=0.012$~GeV$^4$, respectively.
The sum rule in (\ref{eq:SRrhoL}) was found to be quite stable
in a wide range of the threshold parameter $s_0$, allowing its
variation around the central value
$s_0=(m_{\rho'}^2+m_{\rho''}^2)/2$
within the admissible interval of masses
$[m_{\rho'}^2,m_{\rho''}^2]$.
The central values of the decay constant and the moments of the
$\rho'$-meson depend stronger on the threshold $s_0$ than the
$\rho$-meson ground-state values.
This leads to a significant increase of the uncertainties of the
characteristics of the excited $\rho'$-meson state ---
see Table~\ref{tab:rhoL}.
This table collects the main results of our analysis for
$\rho_{\rm L}$, $\rho_{\rm L}^{\prime}$, and $\rho_{\rm T}$
in terms of the meson decay constants, the first five moments,
and the inverse moment of the corresponding DAs.
The first two Gegenbauer coefficients of the
$\rho_{\rm L}$ and the $\rho_{\rm L'}$ DA are given in
Table \ref{tab:rhoLT}.
In both tables, we also include the previous estimates
for $\rho_{\rm L}$ from \cite{BM98} and the corresponding values
of the asymptotic DA (abbreviated by Asy), as well as the
BMS pion DA derived in \cite{BMS01}.
The displayed errors in the parenthesis denote the sum of the
uncertainties related to the variation of the threshold parameter
$s_0$, the normalization constants $f_\rho$ and $f_\rho'$, and the
maximal deviation from the average value of the Borel parameter
$M^2$ in its fiducial window: $M^2\in[0.6, 2]$~GeV$^2$.
Note in passing that an analogous analysis for the pion was
carried out in \cite{BMS01} providing reliable results.
An obvious observation from Table~\ref{tab:rhoL} is that
the new estimates for $\rho_{\rm L}$ and the older results from
\cite{BM98} are more or less compatible to each other.
In contrast, our new estimates for $\rho_{\rm T}$
significantly differ from those for the $\rho_{\rm L}$ DA.
They also disagree with those derived in previous QCD SR
approaches
\cite{CZ84,BB96,BM00,BM01,BJ07},
or were obtained by other ways in \cite{CJ07,D06}.
It is worth mentioning in this context that also the $x$ behavior of
the valence parton distribution functions for $\rho_{\rm T}$
and $\rho_{\rm L}$, extracted from QCD SR calculations in
\cite{Ioffe:2000zd,Oganesian:2001rf}, turns out to be quite different.
%%%%%%%%%%%%%%%%%%%%%%%%%%%%%%%%%%%%%%%%%%%%%%%%%%%%%%%%%%%%%%%%%%%%%%% Table 2
\begin{table}[tbh!]
\caption{Gegenbauer coefficients and slope $\varphi'(0)$ of various
$\rho$-meson DAs, determined at the scale $\mu^2 \approx 1$~GeV$^2$
with QCD SRs employing nonlocal condensates, cf.\ (\ref{eq:SRrhoL}).
In all entries we use $a_6=0$.
The values of the inverse moment $\langle x^{-1}\rangle_\text{model}$
here, in contrast to Table \ref{tab:rhoL}, were computed
with the aid of DA models based on expansions over two Gegenbauer
coefficients.
}
\centering
\label{tab:rhoLT}
 \begin{tabular}{ccccccc}
\hline\noalign{\smallskip}
  Model DA         &$f_M$       &$a_2$       &$a_4$      &$\varphi'(0)$ & $\langle x^{-1}\rangle_\text{model}$
\\[3pt]
\tableheadseprule\noalign{\smallskip}
    Asy              & 1          & 0         & 0           & 6            & 3     \\
$\pi$~\cite{BMS01}   & 0.137(8)   & 0.187(60) & -0.129(40)  & $ 2\pm 6$    & 3.2(1) \\
$\rho_L$~\cite{BM98} & 0.201(5)   & 0.079(20) & -0.074(60)  & $ 2\pm 5$    & 3.0(1)  \\
$\rho_L$  (here)     & 0.21(2)    & 0.047(58) & -0.057(118) & $ 3\pm 8$    & 3.0(2)   \\
$\rho'_L$ (here)     & 0.169(15)  & 0.21(19)  &  0.33(39)   &  --          & --        \\
\noalign{\smallskip}\hline
\end{tabular}
\end{table}
%%%%%%%%%%%%%%%%%%%%%%%%%%%%%%%%%%%%%%%%%%%%%%%%%%%%%%%%%%%%%%%%%%%%%%%

Relying for simplicity only upon the first two Gegenbauer harmonics,
we attempt to model the $\varphi^\text{L}_{\rho}(x,\mu^2)$ meson DA,
in terms of the Gegenbauer coefficients
$a_2=0.047 \pm 0.058$ and $a_4=-0.057 \pm 0.118$
which we determine by means of the first three moments given in
Table \ref{tab:rhoL} at the scale $\mu^2 \approx 1$~GeV$^2$.
The results are given in Table \ref{tab:rhoLT} together with
the coefficients of the DA of the $\rho'$ meson determined in this work.
For the sake of comparison, we also display the results obtained with
the Bakulev--Mikhailov--Stefans (BMS) $\pi$ DA \cite{BMS01}.
Inspection of the displayed values shows that these simple models for
the pion and the $\rho_{\rm L}$ meson are in good mutual agreement, thus
reproducing the observations we made above with respect to the analysis
of the QCD sum rules --- cf.\ Table \ref{tab:rhoL}.

\begin{figure}[t!]
\centering
  \includegraphics[width=0.45\textwidth]{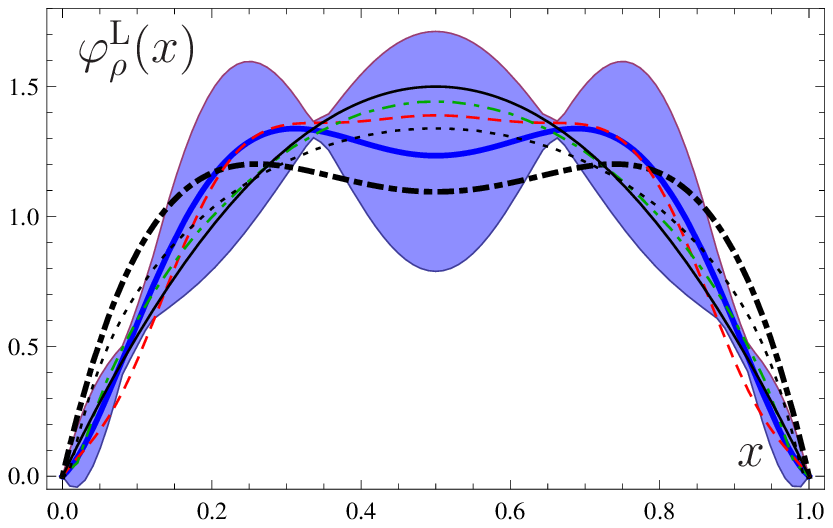}
  \hfill
  \includegraphics[width=0.45\textwidth]{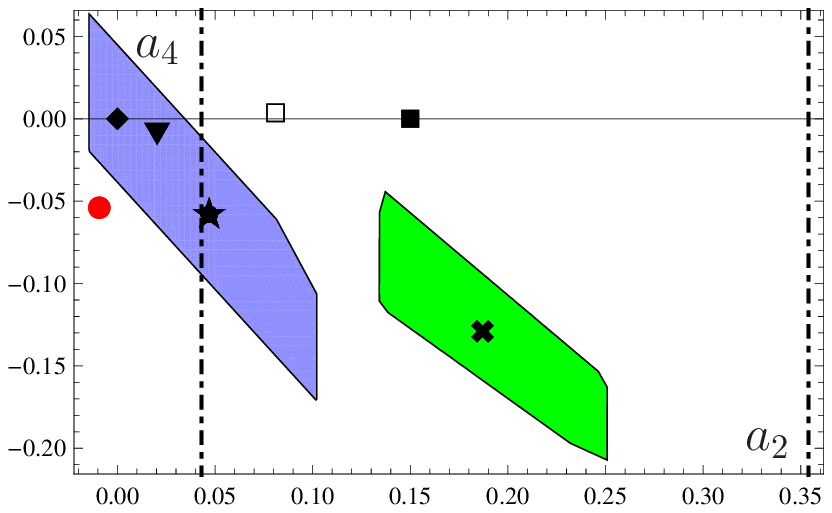}
\caption{\label{fig:rhoL}
 Model distribution amplitudes for the longitudinal $\rho$-meson
 (left panel) and their representations in the plane of the
 Gegenbauer coefficients $a_2$ and $a_4$ (right panel).
 The (blue) shaded areas in both panels indicate the regions of
 the shape variation admitted by our present analysis based on QCD SRs
 with nonlocal condensates.
 The thick solid line inside the DA band corresponds
 to our optimum $\rho_{\rm L}$ DA marked by a $\bigstar$ in the
 $a_2, a_4$ plane.
 The designations for the other DAs in the right panel
 are as follows:
 asymptotic DA
 (solid thin line, \ding{117}),
 QCD SRs~\cite{BJ07}
 (thick dashed-dotted line, \ding{110}),
 AdS QCD~\cite{AhSa12}
 (dotted line, $\square$),
 light-front quark model~\cite{CJ07}
 (thin dashed-dotted line, \ding{116}),
 instanton liquid model~\cite{D06}
 (red dashed line, \mRed{\ding{108}}), and
 lattice QCD~\cite{Lat10}
 (shown only on the right panel by two vertical broken lines) that
 indicate the constraints obtained for $a_2$.
 For comparison, we have displayed in the right panel also
 the area of $a_2, a_4$ values (green slanted rectangle) determined
 with QCD SR and NLC in \cite{BMS01} for the pion DA.
 The BMS model is denoted by the symbol {\footnotesize\ding{54}} ---
 see for details \cite{BMS01,BMS05lat}.
 All models shown were evolved to $\mu^2=1$~GeV$^2$, if they were
 originally determined at another scale.
 }
\end{figure}

The evaluation of the SR in (\ref{eq:SRrhoL}) leads to a whole
set of admissible $\rho_{\rm L}$ DAs, as illustrated in the
left panel of Fig.~\ref{fig:rhoL} in terms of a (blue) shaded ``band''.
To anticipate the particular characteristics of these DAs, we also show
the profiles of other models, proposed in the literature, with further
explanations provided in the figure caption.
The corresponding pairs of Gegenbauer coefficients $a_2, a_4$
of these model DAs, are displayed in the right panel of this figure.
In the plane spanned by the coefficients $a_2$ and $a_4$, the
admissible pairs of values, determined in our analysis, appear in the
form of a slanted (blue) rectangle, within which the symbol $\bigstar$
marks the position of the optimum values in satisfying the SR.
This particular $\rho_{\rm L}$ DA corresponds to the (blue) solid line
within the shaded band on the left.
The designations of the symbols in the right panel of this figure are
defined in the figure caption.
For the sake of comparison, we also show the admissible $a_2, a_4$ set
for the pion determined in \cite{BMS01}
(green shaded area further to the right).
One spots immediately some important differences between the
$\pi$ and the $\rho_{\rm L}$ mesons: First, the confidence
region for the longitudinal $\rho$-meson DA, i.e., the (blue) slanted
rectangle, is located closer to the origin (and the asymptotic DA),
while, second, $a_4$ can have positive values as well.

%%%%%%%%%%%%%%%%%%%%%%%%%%%%%%%%%%%%%%%%%%%%%%%%%%%%%%%%%%%%%%%%%%%%%%% Figure 1
%%%%%%%%%%%%%%%%%%%%%%%%%%%%%%%%%%%%%%%%%%%%%%%%%%%%%%%%%%%%%%%%%%%%%%%

It is instructive to compare the results for the
$\varphi_{\rho}^\text{L}(x,\mu^2)$
DA with their counterparts for $\varphi_\pi(x,\mu^2)$
in more detail, the goal being to clarify the role of the
four-quark contribution $\Phi_\text{4Q}$ in the evaluation of the
SRs given by (\ref{eq:SRrhoL}).
The point is that just this condensate contribution enters
both SRs, those for the pion and those for the $\rho$
(see Eq.\ (\ref{eq:SRrhoL})), but with opposite signs.
In the $\rho$ case, it reduces the total condensate contribution to the
SR because it has the opposite sign relative to the other terms.
No such cancellation occurs in the pion case.
Therefore, the relative weight of $\Phi_\text{4Q}$ in the SR for the
pion DA appears to be enhanced relative to that for the $\rho$.
This entails an increase of the $\varphi_{\pi}$ moments
$
 \va{\xi^{2N}}_{\pi} \geq \va{\xi^{2N}}_{\rho_{\rm L}}
\geq
 \va{\xi^{2N}}_{\rm Asy}
$,
as one sees from Table \ref{tab:rhoL}.
Although the difference of the moments
$\va{\xi^{6}}_{\pi}-\va{\xi^{6}}_{\rm Asy}$
is not very significant within the error bars, leading in turn to a
comparatively small contribution of the Gegenbauer coefficient
$a_{6}^{\pi} \approx 0.059 \sim 0$ , the associated uncertainties
$a_{6}^{\pi}\in[-0.297,0.414]$ are quite large
and comparable in magnitude with the result for the
rho-meson: $a_{6}^{\rho_{\rm L}}=0.05(73)$.

%\vspace*{-10mm}
%%%%%%%%%%%%%%%%%%%%%%%%%%%%%%%%%%%%%%%%%%%%%%%%%%%%%%%%%%%%%%%%%%%%%%% Figure 2
\begin{figure}[h]
%\centering\vspace*{3mm} \vspace*{10mm}
 %\centerline{
 \centering
 \includegraphics[width=0.45\textwidth]{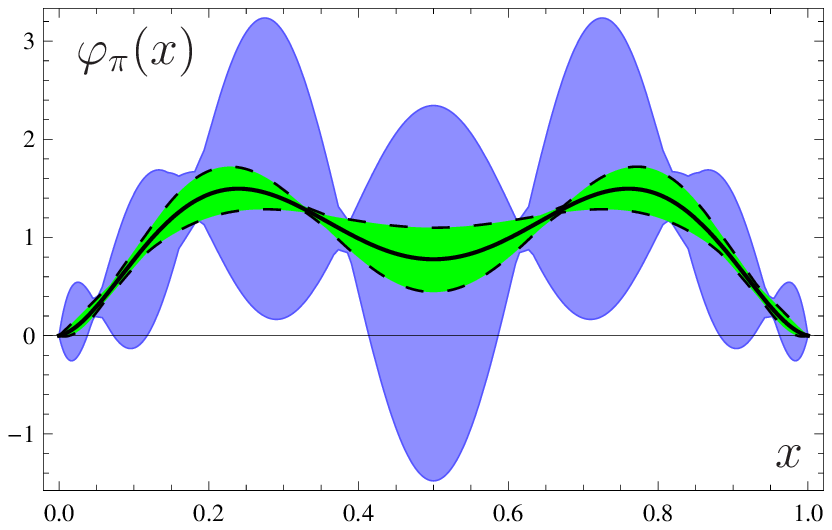}
 \hfill
 \includegraphics[trim = 0mm 15mm 0mm 17mm,clip,width=0.43\textwidth]{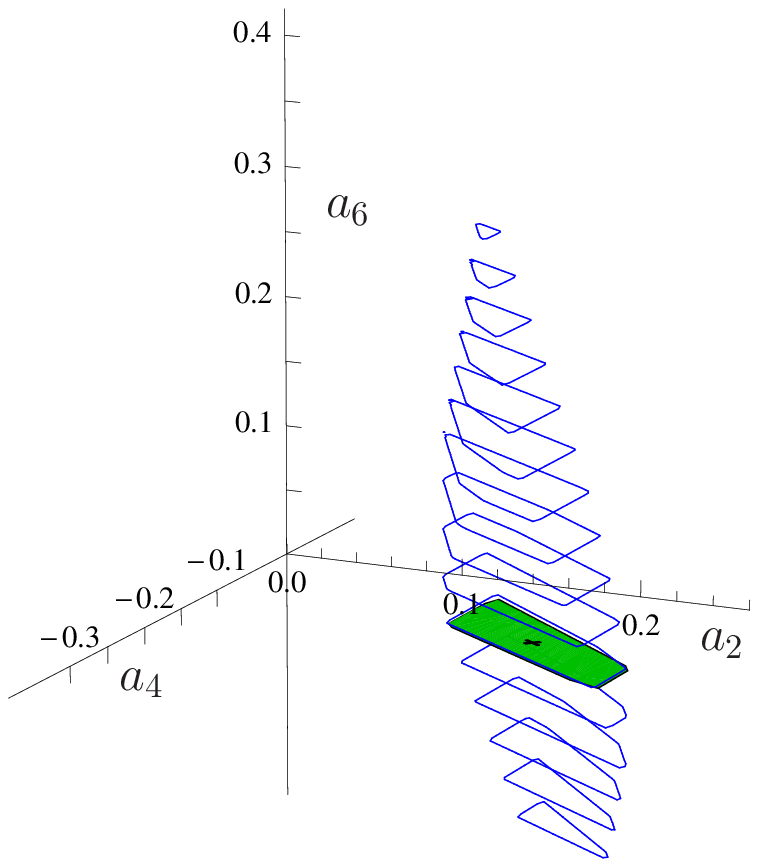}
\vspace*{3mm}
\caption{\label{fig:pi-3D}
Left panel:
  Variation of the pion DA profiles for the 3D
  (blue oscillating wide band) and 2D (green narrow strip) models.
Right panel:
  3D graphics showing the pion DA ``bunch'' \cite{SBMP12},
  obtained from QCD SRs with NLCs, in terms of the coefficients
  $a_2, a_4, a_6$,
  shown as a flight of ``stairs'' of slanted rectangles,
  while the original 2D BMS ``bunch'' \cite{BMS01} in the
  plane $(a_2,a_4)$ is shown as a (green) shaded area, with
  the symbol {\footnotesize\ding{54}} marking the BMS pion DA.
  All results have been obtained at the scale $\mu^2=1$~GeV$^2$.
  }
\end{figure}
%%%%%%%%%%%%%%%%%%%%%%%%%%%%%%%%%%%%%%%%%%%%%%%%%%%%%%%%%%%%%%%%%%%%%%%

However, the inclusion of the $a_6$ uncertainties in the model
for the longitudinal rho meson is not improving its quality owing to
the fact that the values of the lower coefficients $a_2$ and $a_4$,
within their region of validity, are already compatible with the
asymptotic values.
In contrast, in the pion case, one may try to extend the evaluation
of the SRs in such a way as to include the effects of these large
$a_6$ uncertainties, amounting to a 3D analysis \cite{SBMP12}.
This is illustrated in Fig.~\ref{fig:pi-3D}.
In the left panel, we show the profiles of the associated pion DAs
expanded over $a_2, a_4, a_6$  in the form of a 3D (blue) oscillating
band bounded by solid lines in comparison with the original 2D band
(narrower green strip bounded by broken lines) with the value $a_6=0$
determined in \cite{BMS01}.
The inclusion of $a_6$ transforms the original 2D (green) slanted
rectangle into a flight of ``stairs'' of (blue) slanted rectangles
along the $a_6$ axis (right panel).

\section{Conclusions}
\label{sec:Conclusions}
We derived the profile of the leading-twist DA of the longitudinal
$\rho$ meson using QCD sum rules with nonlocal condensates along the
lines of the analysis in \cite{BMS01} for the pion.
We found that the longitudinal $\rho$-meson DA has a shape close to
that obtained in the light-front quark model \cite{CJ07}, bearing
also a resemblance to the profile of a DA derived from the
instanton liquid model \cite{D06}.
From the point of view of the DA key characteristics, expressed
via the inverse moment $\langle x^{-1}\rangle_\text{model}$, we found
that the result obtained with QCD SRs with nonlocal condensates
(Table \ref{tab:rhoL}) agrees, within the determined uncertainties,
reasonably well with the value obtained with a model DA based on the
first two Gegenbauer coefficients (Table \ref{tab:rhoLT}).
We also presented preliminary results on the transverse part of the
$\rho$-meson, notably its decay constant and its moments
$\langle\xi^2\rangle$ and $\langle\xi^4\rangle$, using an improved sum
rule relative to what was considered in \cite{BM00,BM01}.
In contrast to the sum rules employed in \cite{CZ84,BB96}, our sum rule
receives no contributions from the $b_1$-meson term or
$\rho$-meson DAs of higher-twist.
A full-fledged analysis of the $\rho_T$ properties will be given
elsewhere.

\begin{acknowledgements}
The work of A.V.P. was supported by HadronPhysics2,
Spanish Ministerio de Economia y Competitividad
and EU FEDER under contract FPA2010-21750-C02-01, AIC10-D-000598, a-nd
GVPrometeo2009/129.
A.V.P. thanks the organizers of the Light-Cone conference 2013 for
financial support.
We acknowledge support from the Heisenberg--Landau Program under
Grant 2013 and the Russian Foundation for Fundamental
Research (Grants No.\ 12-02-00613a and 11-01-00182a).
\end{acknowledgements}

% BibTeX users please use
%\bibliographystyle{spbasic}
%\bibliographystyle{spbasic4FBS}
%\bibliography{pion,lambda,nonloc,Add4rho}

\begin{thebibliography}{29}
\providecommand{\natexlab}[1]{#1}
\providecommand{\url}[1]{{#1}}
\providecommand{\urlprefix}{URL }
\expandafter\ifx\csname urlstyle\endcsname\relax
  \providecommand{\doi}[1]{DOI~\discretionary{}{}{}#1}\else
  \providecommand{\doi}{DOI~\discretionary{}{}{}\begingroup
  \urlstyle{rm}\Url}\fi
\providecommand{\eprint}[2][]{\url{#2}}

\bibitem[{Ball and Braun(1997)}]{BB97}
Ball, P., Braun, V.M.: {Use and misuse of QCD sum rules in heavy to light
  transitions: The Decay B --- rho e neutrino reexamined}.  Phys. Rev.
  D \textbf{55}, 5561 (1997)
%%CITATION = HEP-PH/9701238;%%

\bibitem[{Ahmady and Sandapen(2013)}]{AhSa12}
Ahmady, M., Sandapen, R.: {Predicting $\bar{B}^0 \to \rho^0 \gamma$ and
  $\bar{B_{s}}^0 \to \rho° \gamma$ using holographic AdS/QCD Distribution
  Amplitudes for the $\rho$ meson}.  Phys. Rev. D \textbf{87}, 054013 (2013)
%%CITATION = ARXIV:1212.4074;%%

\bibitem[{Forshaw and Sandapen(2012)}]{FS12}
Forshaw, J., Sandapen, R.: {An AdS/QCD holographic wavefunction for the rho
  meson and diffractive rho meson electroproduction}.  Phys. Rev. Lett.
  \textbf{109}, 081601 (2012)
%%CITATION = ARXIV:1203.6088;%%

\bibitem[{Efremov and Radyushkin(1980)}]{ER80}
Efremov, A.V., Radyushkin, A.V.: {Factorization and asymptotic behaviour of
  pion form factor in QCD}.  Phys. Lett. B \textbf{94}, 245 (1980)
%%CITATION = PHLTA,B94,245;%%

\bibitem[{Lepage and Brodsky(1980)}]{LB80}
Lepage, G.P., Brodsky, S.J.: {Exclusive processes in perturbative quantum
  chromodynamics}.  Phys. Rev. D \textbf{22}, 2157 (1980)
%%CITATION = PHRVA,D22,2157;%%

\bibitem[{Mikhailov and Radyushkin(1986)}]{MR86ev}
Mikhailov, S.V., Radyushkin, A.V.: {Structure of two loop evolution kernels and
  evolution of the pion wave function in $\phi^3$ in six-dimensions and QCD}.
  Nucl. Phys. B \textbf{273}, 297 (1986)
%%CITATION = NUPHA,B273,297;%%

\bibitem[{Chernyak and Zhitnitsky(1984)}]{CZ84}
Chernyak, V.L., Zhitnitsky, A.R.: {Asymptotic behavior of exclusive processes
  in QCD}.  Phys. Rept. \textbf{112}, 173 (1984)
%%CITATION = PRPLC,112,173;%%

\bibitem[{Ball and Braun(1996)}]{BB96}
Ball, P., Braun, V.M.: {The $\rho$ Meson Light-Cone Distribution Amplitudes of
  Leading Twist Revisited}.  Phys. Rev. D \textbf{54}, 2182 (1996)
%%CITATION = HEP-PH 9602323;%%

\bibitem[{Bakulev and Mikhailov(1998)}]{BM98}
Bakulev, A.P., Mikhailov, S.V.: {The $\rho$-meson and related meson wave
  functions in QCD sum rules with nonlocal condensates}.  Phys. Lett.
  B \textbf{436}, 351 (1998)
%%CITATION = PHLTA,B436,351;%%

\bibitem[{Bakulev and Mikhailov(2000)}]{BM00}
Bakulev, A.P., Mikhailov, S.V.: {QCD vacuum tensor susceptibility and
  properties of transversely polarized mesons}.  Eur. Phys. J. C \textbf{17},
  129 (2000)
%%CITATION = HEP-PH 9908287;%%

\bibitem[{Bakulev and Mikhailov(2001)}]{BM01}
Bakulev, A.P., Mikhailov, S.V.: {New shapes of light-cone distributions of the
  transversely polarized rho mesons}.  Eur. Phys. J. C \textbf{19}, 361   (2001)
%%CITATION = HEP-PH 0006206;%%

\bibitem[{Arthur  et~al(2011)}]{Lat10}
Arthur, R., et~al: {Lattice Results for Low Moments of Light Meson Distribution
  Amplitudes}.  Phys. Rev. D \textbf{83}, 074505 (2011)
%%CITATION = 1011.5906;%%

\bibitem[{de~Teramond and Brodsky(2009)}]{TB08}
de~Teramond, G.F., Brodsky, S.J.: {Light-Front Holography: A First
  Approximation to QCD}.  Phys. Rev. Lett. \textbf{102}, 081601 (2009)
%%CITATION = ARXIV:0809.4899;%%

\bibitem[{Vega  et~al(2009)Vega, Schmidt, Branz, Gutsche, and
  Lyubovitskij}]{VSBG09}
Vega, A., Schmidt, I., Branz, T., Gutsche, T., Lyubovitskij, V.E.: {Meson wave
  function from holographic models}.  Phys. Rev. D \textbf{80}, 055014 (2009)
%%CITATION = ARXIV:0906.1220;%%

\bibitem[{Choi and Ji(2007)}]{CJ07}
Choi, H.M., Ji, C.R.: {Distribution amplitudes and decay constants for (pi, K,
  rho, K*) mesons in light-front quark model}.  Phys. Rev. D \textbf{75}, 034019
  (2007)
%%CITATION = HEP-PH/0701177;%%

\bibitem[{Dorokhov(2006)}]{D06}
Dorokhov, A.E.: {Distribution amplitudes of light mesons and photon in the
  instanton model}.  Czech. J. Phys. \textbf{56}, F169 (2006); Braz. J. Phys. \textbf{37}, 819 (2007)
%%CITATION = HEP-PH/0610212;%%

\bibitem[{Mikhailov and Radyushkin(1989)}]{MR89}
Mikhailov, S.V., Radyushkin, A.V.: {Quark condensate nonlocality and pion wave
  function in QCD}.  Sov. J. Nucl. Phys. \textbf{49}, 494 (1989)
%%CITATION = SJNCA,49,494;%%

\bibitem[{Mikhailov and Radyushkin(1992)}]{MR92}
Mikhailov, S.V., Radyushkin, A.V.: {The pion wave function and QCD sum rules
  with nonlocal condensates}.  Phys. Rev. D \textbf{45}, 1754 (1992)
%%CITATION = PHRVA,D45,1754;%%

\bibitem[{Bakulev and Mikhailov(2002)}]{BM02}
Bakulev, A.P., Mikhailov, S.V.: {Lattice measurements of nonlocal quark
  condensates, vacuum correlation length, and pion distribution amplitude in
  QCD}.  Phys. Rev. D \textbf{65}, 114511 (2002)
%%CITATION = HEP-PH 0203046;%%

\bibitem[{Bakulev  et~al(2004)Bakulev, Mikhailov, and Stefanis}]{BMS03}
Bakulev, A.P., Mikhailov, S.V., Stefanis, N.G.: {CLEO and E791 data: A smoking
  gun for the pion distribution amplitude?}  Phys. Lett. B \textbf{578}, 91   (2004)
%%CITATION = HEP-PH 0303039;%%

\bibitem[{Mikhailov  et~al(2010)Mikhailov, Pimikov, and Stefanis}]{MPS10}
Mikhailov, S.V., Pimikov, A.V., Stefanis, N.G.: {Endpoint behavior of the pion
  distribution amplitude in QCD sum rules with nonlocal condensates}.  Phys.
  Rev. D \textbf{82}, 054020 (2010)
%%CITATION = 1006.2936;%%

\bibitem[{Bakulev  et~al(2001)Bakulev, Mikhailov, and Stefanis}]{BMS01}
Bakulev, A.P., Mikhailov, S.V., Stefanis, N.G.: {QCD-based pion distribution
  amplitudes confronting experimental data}.  Phys. Lett. B \textbf{508},
  279 (2001); Erratum: ibid. B \textbf{590}, 309 (2006)
%%CITATION = HEP-PH 0103119;%%

\bibitem[{Beringer  et~al(2012)}]{PDG12}
Beringer, J., et~al: {Review of Particle Physics (RPP)}.  Phys. Rev.
  D \textbf{86}, 010001 (2012)
%%CITATION = PHRVA,D86,010001;%%

\bibitem[{Bakulev  et~al(2003)Bakulev, Mikhailov, and Stefanis}]{BMS02}
Bakulev, A.P., Mikhailov, S.V., Stefanis, N.G.: {Unbiased analysis of {CLEO}
  data beyond {LO} and pion distribution amplitude}.  Phys. Rev. D \textbf{67},
  074012 (2003)
%%CITATION = HEP-PH 0212250;%%

\bibitem[{Ball and Jones(2007)}]{BJ07}
Ball, P., Jones, G.: {Twist-3 distribution amplitudes of K* and phi mesons}.
  JHEP \textbf{0703}, 069 (2007)
%%CITATION = HEP-PH/0702100;%%

\bibitem[{Ioffe and Oganesian(2001)}]{Ioffe:2000zd}
Ioffe, B., Oganesian, A.: {Valence quark distributions in mesons in generalized
  QCD sum rules}.  Phys. Rev. D \textbf{63}, 096006 (2001)
%%CITATION = HEP-PH/0011348;%%

\bibitem[{Oganesian and Samsonov(2001)}]{Oganesian:2001rf}
Oganesian, A., Samsonov, A.: {Second moment of quark structure function of the
  rho meson in QCD sum rules}.  JHEP \textbf{0109}, 002 (2001)
%%CITATION = HEP-PH/0107077;%%

\bibitem[{Bakulev  et~al(2006)Bakulev, Mikhailov, and Stefanis}]{BMS05lat}
Bakulev, A.P., Mikhailov, S.V., Stefanis, N.G.: {Tagging the pion quark
  structure in QCD}.  Phys. Rev. D \textbf{73}, 056002 (2006)
%%CITATION = HEP-PH 0512119;%%

\bibitem[{Stefanis  et~al(2013)Stefanis, Bakulev, Mikhailov, and
  Pimikov}]{SBMP12}
Stefanis, N.G., Bakulev, A.P., Mikhailov, S.V., Pimikov, A.V.: {Can We
  Understand an Auxetic Pion-Photon Transition Form Factor within QCD?}
  Phys. Rev. D \textbf{87}, 094025 (2013)
%%CITATION = ARXIV:1202.1781;%%

\end{thebibliography}

%\end{document}

\newcommand{\noopsort}[1]{} \newcommand{\printfirst}[2]{#1}
  \newcommand{\singleletter}[1]{#1} \newcommand{\switchargs}[2]{#2#1}

\end{document}